\documentclass[10pt,letterpaper,twocolumn]{article} 

\usepackage{ol2}
\usepackage[draft]{hyperref}
\usepackage{amsmath}
\usepackage{cite}
\usepackage{graphicx}

\begin{document}
\twocolumn[
\title{Multiple Side-Band Generation for Two-Frequency Components Injected into a Tapered Amplifier}
\author{Hua Luo,$^1$ Kai Li,$^1$ Dongfang Zhang,$^1$ Tianyou Gao,$^1$ and Kaijun Jiang$^{1,2,*}$}
\address{
$^1$State Key Laboratory of Magnetic Resonance and Atomic and
Molecular Physics, Wuhan Institute of Physics and Mathematics,
Chinese Academy of Sciences, Wuhan, 430071, China\\$^2$Center for
Cold Atom Physics, Chinese Academy of Sciences, Wuhan, 430071,
China\\$^*$Corresponding Author: kjjiang@wipm.ac.cn}

\begin{abstract}
We have experimentally studied the multiple side-band generation
for two-frequency components injected into a tapered amplifier
and demonstrated its effects on atomic laser cooling. A heterodyne
frequency-beat measurement and a Fabry Perot interferometer have
been applied to analyze the side-band generation with different
experimental parameters, such as frequency difference, injection
laser power and tapered amplifier current. In laser cooling
potassium40 and potassium41 with hyperfine splitting of 1.3GHz and
254MHz, respectively, the side-band generation with a small frequency difference has a significant
effect on the number of trapped atoms.

\end{abstract}
\ocis{140.3520, 140.4480, 190.4223, 020.3320}]

\noindent Multiple side bands with an equal frequency spacing have
various applications in quantum optics and laser cooling. Zhu {\it
et. al.} observed the multiple-mode side-band fluorescence around
the resonant transition with a strong dichromatic field
coupling two energy levels \cite{yifu1990twofreq}. Multiple electromagnetic induced transparencies
\cite{wangjin2003biEIT} and four-wave mixing have been observed with
two coupling components and one probing component in a three-level
system \cite{zhan2010FWM}. And an arbitrary side-band with squeezing
has been demonstrated in the experiment \cite{kozuma2010sqeezing}.
In laser cooling of alkali atoms, two frequency components are required, one
for optical pumping and the other for cooling. The semiconductor
laser has become a standard tool in laser cooling due to its easy
operation and low price. But its low output power (generally less
than 100mW) puts a limit on the number of trapped atoms. The output
power of a tapered amplifier can be very high (generally more than
500mW), which enables us to improve the parameters of trapped atoms. Different
spectra have been achieved on the base of a tapered amplifier. For
example, two-frequency components can be simultaneously injected
into a tapered amplifier \cite{Salomon1999twofreq}. A high power
output around 670nm has been obtained with an external coupling
cavity \cite{Tien2008cavitytaped, Chi2010cavitytaped}. And a tapered
amplifier has been injection seeded with a femtosecond comb
\cite{Ye2007fstaped}.

To get a high-power multi-mode output, multiple frequency components can be simultaneously amplified in a tapered amplifier to simplify the experimental setup. But at the same time, the wave-mixing process can result in a multiple side-band generation, which may have non-negligible effects on atomic behaviors. Here, we will use a heterodyne frequency-beat measurement with a high resolution ($<$500kHz) and a FPI (Fabry Perot interferometer) to analyze the side-band generation when two-frequency components are injected into a tapered
amplifier. Laser cooling potassium40 and potassium41 with different hyperfine splitting can demonstrate the
effect of the side-band generation on the interaction between atoms and light.

Our experimental setup is shown in Fig.1. Each half wave plate (from HWP1 to HWP8) is used to rotate the laser polarization direction for adjusting the relative powers in the two outputs of its following PBS (polarization beam splitter). The output of a diode laser (767nm wavelength, 30mW output power and a vertical polarization) is divided into two beams in PBS1. The reflected beam with a vertical polarization is frequency shifted by AOM1 (acousto-optic modulator) as a reference beam ($\nu_r$) for the heterodyne frequency-beat measurement. The transmitting beam with a horizontal polarization is again divided into two beams in PBS2, and then one beam is frequency shifted from 40MHz to 1GHz by using AOM2. The two beams combine together in PBS3 and then are reflected by PBS4. Before injected into a TA (tapered amplifier), the beam contains two-frequency components ($\nu_1$ and $\nu_2$) both of which have a vertical polarization to match the TA polarization. After the TA, The main power is reflected by PBS5 for laser cooling and a small fraction transmits PBS5 for side-band measurement. In PBS6, the reflected beam with a vertical polarization is analyzed by a FPI (free spectral range 3GHz) and the transmitting one overlaps with the reference beam in PBS7 for heterodyne frequency-beat measurement. After PBS8, the two beams with a horizontal polarization go through a fiber and are detected by a high-speed photodetector (Newport, Model 1343, 25GHz bandwidth). For our used TA (UniQuanta Tech, Model TAL100, spectrum range 750nm-775nm), the output power is 700mW in $21^0$C when the injection power is 14mW and the tapered amplifier current is 1.8A. The apertures of the front and rear facets are 3um and 190um, respectively, and both facets are anti-reflection coated with $0.01\%$. The chip is based on the semiconductor material GaAs and the amplification length is 2.5mm. For our frequency-beat measurement, the two beams coming from one diode laser can produce a stable wave-mixing due to a good phase coherence.

\begin{figure}[htb]
\centerline{\includegraphics[width=8.0cm]{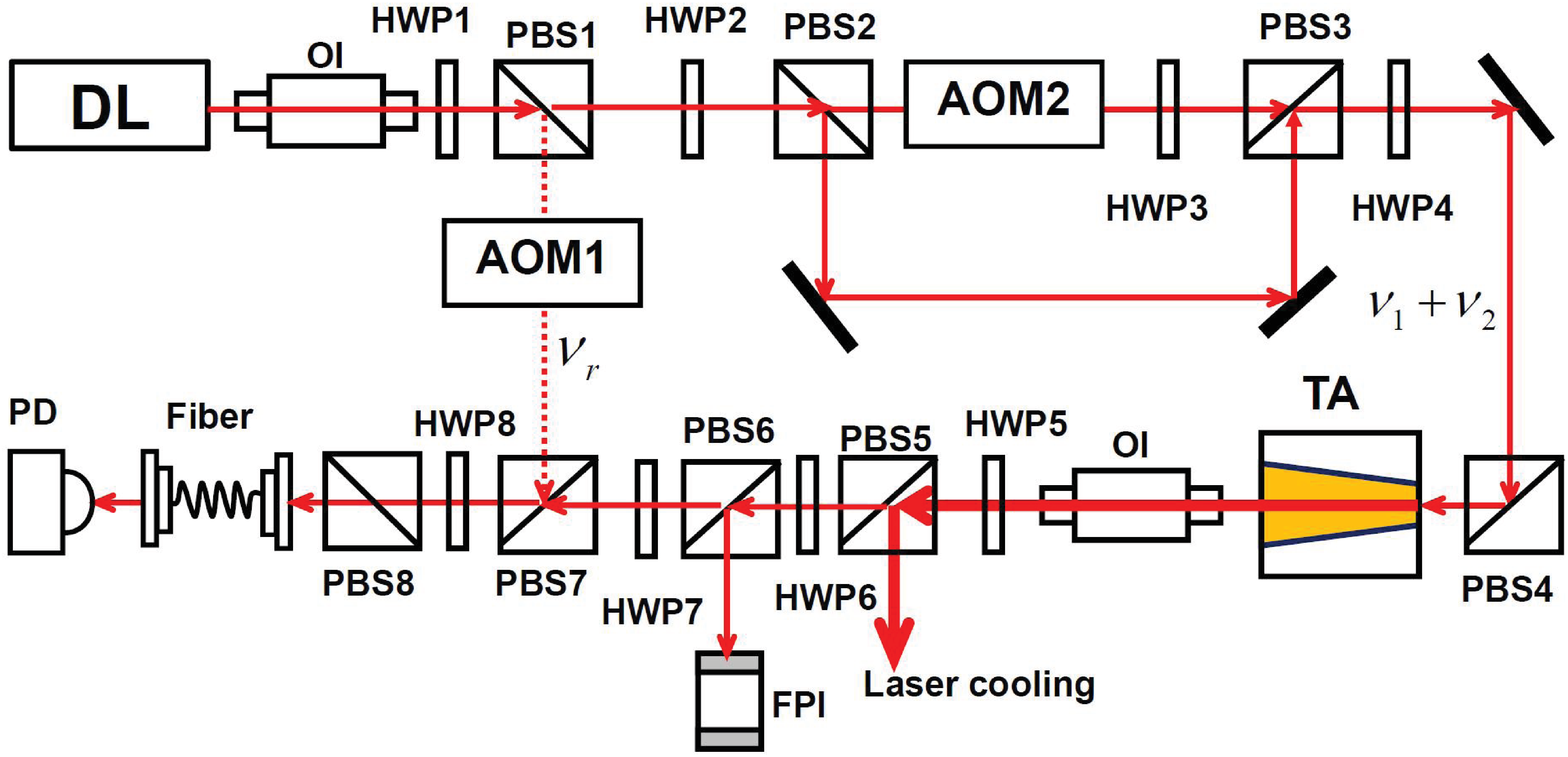}}
\caption{(Color online) Shematics of the experimental setup. DL: diode laser, OI: optical isolator, HWP: half-wave plate,
PBS: polarization beam splitter, AOM: acousto-optic modulator, TA: tapered amplifier, PD: high-speed photodetector, FPI: Fabry Perot
interferometer. The dotted line indicates the reference laser beam.}
\end{figure}

In our experiment, we use a heterodyne measurement to analyze the side-band generation because of the following three reasons: (1) A FPI couldn't distinguish different components for a small frequency difference due to its limited resolution. For our applied FPI, its resolution is about 100MHz. (2) Frequency beat between the high-order side bands is too small to be detected due to its low power distribution. (3) In the homodyne measurement, different pairs of components may contribute to the same frequency-beat peak. While in the heterodyne measurement, different frequency components can be distinguished clearly. In Fig.2(a), the powers of the reference beam and the main beam are 0.67mW and 2.28mW, respectively. The frequency difference between the two injection components is 80MHz, $\delta=\nu_2-\nu_1=80MHz$, and the reference component is between these two components, $\Delta=\nu_2-\nu_r=20MHz$ and $\nu_r-\nu_1=60MHz$. For positive side bands, $f_{+n}=\Delta+n\delta$, and for negative side bands, $f_{-n}=(\delta-\Delta)+n\delta$, where $n$ is the order of the side bands. When $n=0$, the signals correspond to the two injection frequencies ($\nu_2$ and $\nu_1$). All the frequency components (including injection components and side bands) are indicated in the Fig.2(a). Here, we can observe the 6th order side-band generation, which is the highest order that has been experimentally obtained in a TA. There are also homodyne frequency-beat signals between different components in the output of the TA, $f_m=m\delta$, where $m=1,2,3,...p-1$ and $p$ is the total number of the frequency components. Only up to $m=7$ signal can be distinguished in our detection sensitivity and the frequency-beat between higher order side bands are negligible small. We also do the heterodyne frequency-beat for a 200MHz difference, where $\nu_2-\nu_1=200MHz$ and $\nu_2-\nu_r=60MHz$. The frequency-beat signals are indicated in the insert of Fig.2(a) and only 3 side bands (not include the sign of the order) can be detected. When a FPI is applied to perform the measurement, the side-band number is also 3 as shown in Fig.3(a). For frequency differences of 280MHz and 360MHz, the measured numbers of the side-band with these two methods are also the same. As shown in Fig.2(b), the side-band number decreases as the frequency difference increasing.

\begin{figure}[htb]
\centerline{\includegraphics[width=7.5cm]{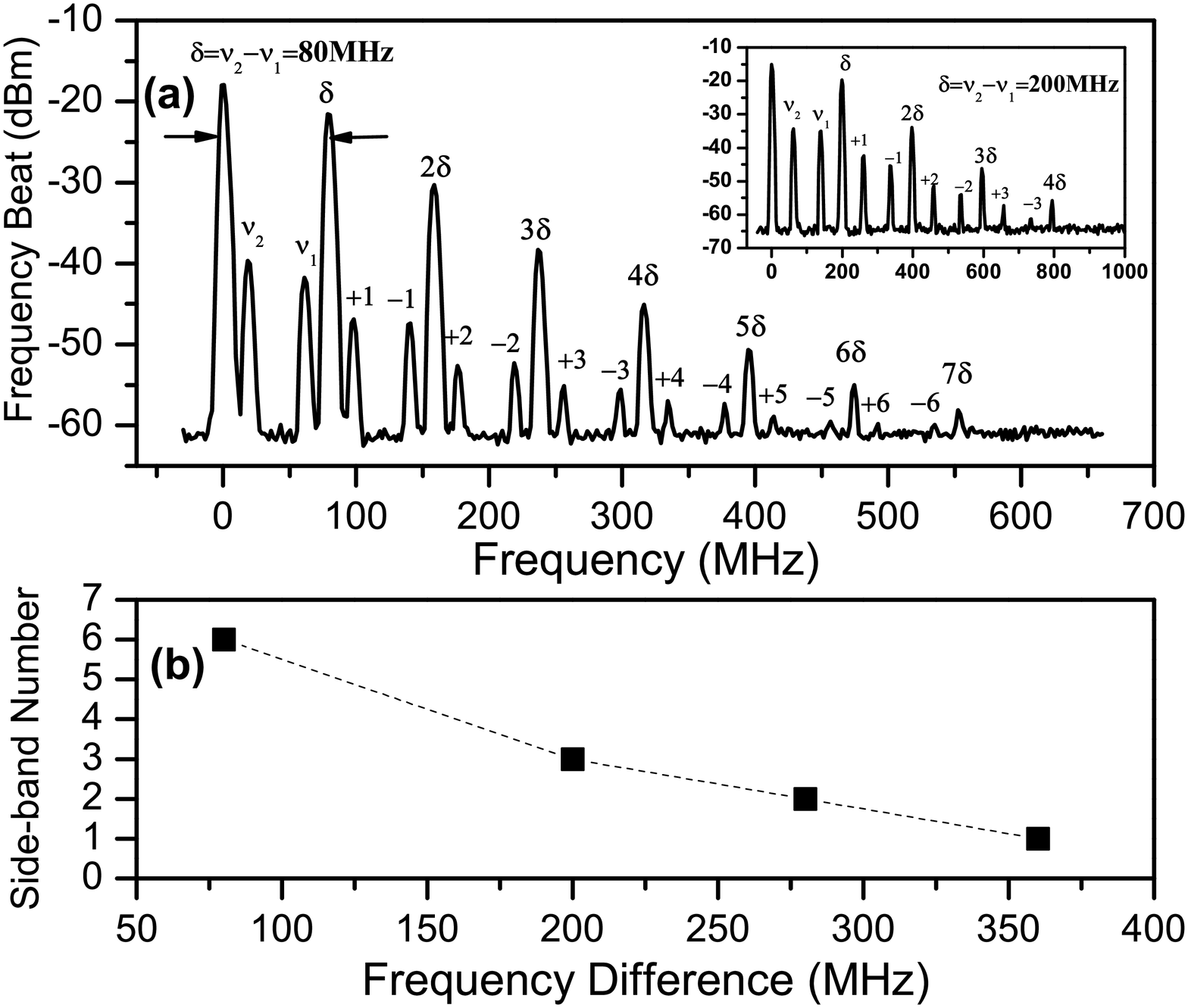}}
\caption{Generated side bands versus the frequency difference, where the injected power of both components is 15mW and the TA current is 1.8A. (a) Frequency beat signal for a 80MHz frequency difference, and the insert is for a 200MHz frequency difference. (b) Side-band number versus the frequency difference.}
\end{figure}

Multiple side bands originate from wave mixing when different
frequency components overlap in a TA. This nonlinear
process results in the laser power distributing among different side
bands. A FPI is applied to measurement the power distribution among different side bands for a 200MHz frequency difference. As shown in Fig.3(a), there exist positive and negative three order side bands. When injection powers
of the two components are equivalent, side bands distribute almost symmetrically on both sides. Throughout this manuscript, the side-band power is denoted by its percentage ratio in the total output for simplification. Since the signal for the third order side band will be indistinguishable in a small TA current, we only include the experimental results for positive and negative two orders. As shown
in Fig.3(b), the power of each side band almost remains unchanged with different TA current.
This implies that the two components undergo wave mixing first to get multiple side bands and then are amplified proportionally. The small power difference of the same order side band (for example $+\delta$ and $-\delta$) mainly comes from the power imbalance between injection components.

Fig.4 shows that the total power in all side bands almost remains unchanged at $23\%$ with the TA current varying as indicated in Fig.3(b). It is also shown in Fig.4 that the total power increases from $14\%$ to $27\%$ with injection laser power increasing. The multiple side-band generation results from a nonlinear process in the solid material and it should be enhanced in the saturation regime. So, the wave-mixing process will become more dominant with a higher injection power.

\begin{figure}[htb]
\centerline{\includegraphics[width=7.5cm]{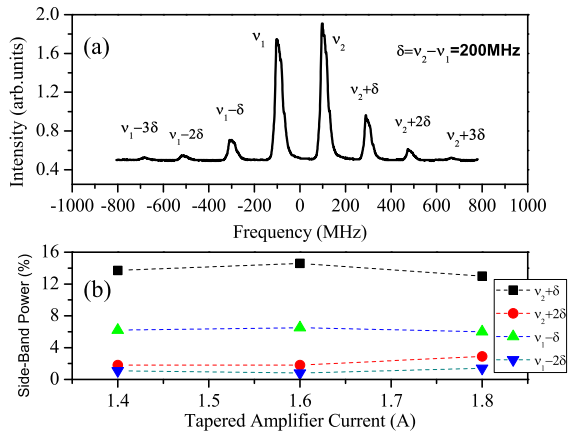}}
\caption{(Color online) Power distribution of each side band with a 15mW injection laser power and a 200MHz frequency difference. (a) Side-band signal measured with a FPI. In a 1.8A tapered amplifier current, up to $+3\delta$ and $-3\delta$ side bands are obvious on the spectrum. (b) Power distribution of each side band. The black square is for $+\delta$, green uptriangle for $-\delta$, red circle for $+2\delta$ and blue downtriangle for $-2\delta$.}
\end{figure}

\begin{figure}[htb]
\centerline{\includegraphics[width=7.0cm]{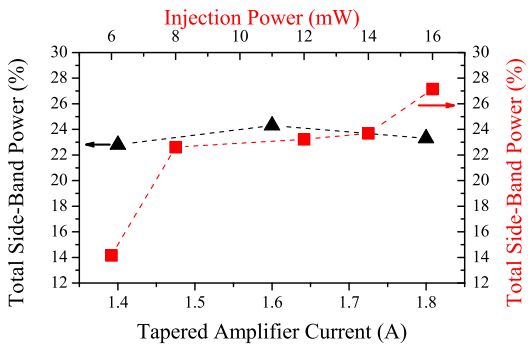}}
\caption{(Color online) Total power of all side bands. The black triangle denotes the total power versus the tapered amplifier
current and the red square denotes that versus the injection power.}
\end{figure}

The multiple side-band generation should have effects on the interaction
between atoms and light. For example in atomic laser cooling, the
side-band generation would decrease the power of the cooling light and additional excitations would destroy trapped atoms. To
verify this prediction, we first do laser cooling of potassium40 with a 1.3GHz hyperfine splitting in the ground state.
Two equivalent frequency components with a 1.3GHz difference are injected into
a TA and only 35mw power of the output is used to cool atoms. The number of trapped atoms is $4.2(1.3)\times10^8$. As a comparison, we directly use two equivalent laser beams with a 1.3GHz frequency difference and a total 35mw laser power to do the experiment. The trapped atom number is $3.7(1.5)\times10^8$ in this condition. The similar numbers of obtained atoms demonstrate no obvious effect of the side band. But for potassium41 with only a 254MHz hyperfine splitting, the trapped atom number is $1.8(0.8)\times10^8$ with a two-frequency
injection and $2.1(1.0)\times10^9$ without the two-component injection. The one order of magnitude difference on the trapped atomic numbers shows a significant effect of the side-band generation. From our measurement, the total side-band power is about
$20\%$ for a 254MHz frequency difference and below our detection sensitivity for a 1.3GHz difference. This study demonstrates that
the two-frequency injection in a TA can afford a convenient way to simplify the experimental setup on one side. But
on the other side, the multiple side-band generation has a notable effect on the atomic behaviors if the frequency difference is
small. Laser cooling potassium40 is our starting point to realize degenerate Fermi gases.

We emphasize that the heterodyne frequency-beat measurement is the first time applied to analyze the side-band generation in a TA. This method is still efficient for a small frequency difference due to its high spectrum resolution. Also, a FPI is applied to
explore the side-band generation with different parameters, such as frequency difference, tapered amplifier current and injection laser power. Injection of two frequency components into a TA has been widely employed in laser cooling (for example the reference \cite{Inguscio2001K}) and this indicates the significance of a TA with a multiple-frequency injection. But authors of all these previous works didn't consider the effect of the side-band generation. In this manuscript, laser cooling potassium40 and potassium41 demonstrates its obvious effect on the interaction between atoms and light.

This work is supported by NSFC (Grant No 11004224) and NFRF-China (Grant No 2011CB921601).


\begin{thebibliography}{1}
\newcommand{\enquote}[1]{``#1''}

\bibitem{yifu1990twofreq}
Y.~F. Zhu, Q.~L. Wu, A.~Lezama, D.~J. Gauthier, and T.~W. Mossberg,
  \enquote{Resonance fluorescence of two-level atoms under strong bichromatic
  excitation,} Phys. Rev. A \textbf{41}, 6574 (1990).

\bibitem{wangjin2003biEIT}
J.~Wang, Y.~F. Zhu, K.~J. Jiang, and M.~S. Zhan, \enquote{Bichromatic
  electromagnetically induced transparency in cold rubidium atoms,} Phys. Rev.
  A \textbf{68}, 063810 (2003).

\bibitem{zhan2010FWM}
G.~Q. Yang, P.~Xu, J.~Wang, Y.~F. Zhu, and M.~S. Zhan, \enquote{Four-wave
  mixing in a three-level bichromatic electromagnetically induced transparency
  system,} Phys. Rev. A \textbf{82}, 045804 (2010).

\bibitem{kozuma2010sqeezing}
M.~Arikawa, K.~Honda, D.~Akamatsu, S.~Nagatsuka, K.~Akiba, A.~Furusawa, and
  M.~Kozuma, \enquote{Quantum memory of a squeezed vacuum for arbitrary
  frequency sidebands,} Phys. Rev. A \textbf{81}, 021605(R) (2010).

\bibitem{Salomon1999twofreq}
G.~Ferrari, M.~Mewes, F.~Schreck, and C.~Salomon, \enquote{High-power
  multiple-frequency narrow-linewidth laser source based on a semiconductor
  tapered amplifier,} Opt. Lett. \textbf{24}, 151 (1999).

\bibitem{Tien2008cavitytaped}
T.~Q. Tien, M.~Maiwald, B.~Sumpf, G.~Erbert, and G.~Tr\"ankle,
  \enquote{Microexternal cavity tapered lasers at 670 nm with 5 w peak power
  and nearly diffraction-limited beam quality,} Opt. Lett. \textbf{33}, 2692
  (2008).

\bibitem{Chi2010cavitytaped}
M.~Chi, G.~Erbert, B.~Sumpf, and P.~M. Petersen, \enquote{Tunable high-power
  narrow-spectrum external-cavity diode laser based on tapered amplifier at
  668nm,} Opt. Lett. \textbf{35}, 1545 (2010).

\bibitem{Ye2007fstaped}
F.~C. Cruz, M.~C. Stowe, and J.~Ye, \enquote{Tapered semiconductor amplifiers
  for optical frequency combs in the near infrared,} Opt. Lett. \textbf{31},
  1337 (2006).

\bibitem{Inguscio2001K}
G.~Modugno, G.~Ferrari, G.~Roati, R.~J. Brecha, A.~Simoni, and M.~Inguscio,
  \enquote{Bose-einstein condensation of potassium atoms by sympathetic
  cooling,} Science \textbf{294}, 1320 (2001).

\end{thebibliography}
\end{document}